# Surface analysis of tiles and samples exposed to the first JET campaigns with the ITER-Like Wall

J P Coad[a,ci], E Alves[b], N P Barradas[c], A Baron-Wiechec[d], K Heinola[e], J Likonen[a], M Mayer[e], G F Matthews[d], P Petersson[g], A Widdowson[d] and JET-EFDA contributors[1]

*JET-EFDA, Culham Science Centre, OX14 3DB, Abingdon, UK*
[a]*Association EURATOM-TEKES, VTT, PO Box 1000, 02044 VTT, Espoo, Finland*
[b]*Associação EURATOM/IST, Instituto de Plasmas e Fusão Nuclear, Universidade de Lisboa, 1049-001 Portugal*
[c]*Associação EURATOM/IST, Centro de Ciências e Tecnologias Nucleares, Universidade de Lisboa, 2686-953 Sacavém, Portugal*
[d]*EURATOM/CCFE Fusion Association, Culham Science Centre, Abingdon, OX14 3DB, UK*
[e]*Association EURATOM-TEKES, University of Helsinki, PO Box 64, 00014 University of Helsinki, Finland*
[f]*Max-Planck-Institut fuer Plasmaphysik, EURATOM Association, Boltzmannstr. 2, 85748 Garching, Germany*
[g]*Department for Fusion Plasma Physics, Association EURATOM-VR, Royal Institute of Technology, S-10405 Stockholm, Sweden*

**Abstract**

This paper reports on the first post-mortem analyses of tiles removed from JET after the first campaigns with the ITER-like Wall (ILW) during 2011-2 [1]. Tiles from the divertor have been analysed by the Ion Beam Analysis (IBA) techniques Rutherford Backscattering Spectroscopy (RBS) and Nuclear Reaction Analysis (NRA) and by Secondary Ion Mass Spectrometry (SIMS) to determine the amount of beryllium deposition and deuterium retention in the tiles exposed to the scrape-off layer. Films 10-20 microns thick were present at the top of Tile 1, but only very thin films (<1 micron) were found in the shadowed areas and on other divertor tiles. The total amount of Be found in the divertor following the ILW campaign was a factor of ~9 less that the material deposited in the 2007-9 carbon campaign, after allowing for the longer operations in 2007-9.

**Introduction**

This paper reports on the first post-mortem analyses of tiles removed from JET after the first ITER-like Wall (ILW) operations in 2011-2 [1], and is the first stage of a comprehensive analysis strategy planned by the JET-EFDA Task Force Fusion Technology (TFFT) [2]. A cross-section of the JET divertor with the tile numbers and the plasma configuration that was used predominantly during ILW operations is shown in Figure 1. In previous JET campaigns with the carbon wall, the main plasma impurity was carbon, with ~10% Be originating from the regular Be evaporations in the main chamber, and ~1% made up of Ni, Fe, Cr and other metals from the inconel vessel walls and tile fixings [3]. This was the basic composition of deposited films in the main chamber, and of the material travelling along the SOL towards the inner divertor to deposit principally on tiles 1 and 3; since its installation in 2006 there has also been some deposition on the High Field Gap Closure (HFGC) tile [4-6]. At tiles 1 and 3 some of the deposits were re-eroded, particularly the carbon, by chemical sputtering by deuterium, with migration to tile 4 and thence to the shadowed regions at the inner divertor corner where co-deposition with deuterium occurred. Tiles 1 and 3 were left with films rich in Be and the other metals (Be/C typically ~1). At the outer divertor there was also deposition on tile 6 and in the shadowed regions: there was normally no residual deposition on tiles 7 and 8, and little on the Load Bearing Tile (LBT) [4-6]. If similar migration occurs with a metal wall, this would pose a problem for ITER, so the initial attention has been paid to those divertor areas where deposition occurred in previous campaigns. Tiles and witness samples have been analysed by Ion Beam Analysis (IBA) and by Secondary Ion Mass Spectrometry (SIMS) to determine the amount and mix of beryllium/carbon deposition and deuterium retention during the ILW campaign. More comprehensive analyses will follow, and other divertor areas will be checked for deposition and/or erosion.

---
[1] See the Appendix of F Romanelli et al., Proc 24[th] IAEA Fusion Energy Conference 2012, San Diego, USA

**Experimental and Results**

As shown in Fig. 1 the outer strike point was usually located on the outer part of the Load Bearing Tile (LBT) and the inner strike point was near the top of Tile 3, so that the inner SOL along which most impurities in JET migrate intersects Tile 1 and the High Field Gap Closure (HFGC) tile - the so-called High Delta or HD-configuration. This differs from previous campaigns with the carbon wall when there were proportionately many more pulses with the inner strike point low on tile 3 or on tile 4 [5].

A full poloidal set of divertor marker tiles were present during operations, and have been removed for analysis. The standard tile is a carbon-fibre reinforced carbon composite (CFC) substrate coated by a Physical Vapour Deposition technique with 20 microns of W on a Mo interfacial layer (to reduce inter-diffusion) [7]. For marker tiles the W coating is stopped after about 18 microns, then ~3 microns Mo and ~4 microns W layers are applied on top of the thick W coating: the position of the Mo interlayer allows a change in thickness of the outer W layer resulting from operations to be determined.

Following the ILW operations in 2011-2 all the divertor tiles appear by eye to be very similar to the freshly W-coated divertor tiles prior to operation in regions that were, or could be, exposed to plasma. Parts of tiles 4 and 6 that were permanently shadowed from direct plasma exposure by tile 3 or tile 7 and other surfaces within shadowed regions were covered with dark, coloured, films.

Divertor tiles 1, 4, 6 and HFGC have been analysed by IBA techniques such as Rutherford Backscattering Spectroscopy (RBS) and Nuclear Reaction Analysis (NRA). The RBS spectra shown here were recorded with a proton ($^1$H) ion beam energy of 2.3 MeV and the usual measurement charge was 1 µC. The Si PIPS detector from Canberra with a nominal resolution of 12 keV and 100 µm active layer was at a scattering angle of 135° subtending a solid angle of 4.25 msr at the target and with no stopping foil. Proton Induced X-ray Emission (PIXE) spectra were recorded simultaneously, but are not shown here. In NRA a $^3$He beam produces the reactions $^9$Be($^3$He,p)$^{11}$B, $^{12}$C($^3$He,p)$^{14}$N and $^2$H($^3$He,p)$^4$He for analysis of Be, C and $^2$H (or D, deuterium) respectively. The beam energy for NRA was also 2.3 MeV, whilst the measuring charge was usually 2 µC. The Ortec Si detector had a nominal resolution of 20 keV and a 1500 µ active layer, and was at a scattering angle of 135° with a 140µm aluminium stopping foil and subtending 19.81 msr at the target. $^3$He-RBS spectra were recorded at the same time as the NRA, but are not included here. The beams were generated using a 2.5 MeV Van de Graaf, and the beam diameter at the target was 1mm. These techniques analyse to a depth of a few microns at the surface, so to assess the profile of the elements within the surface and to monitor the composition to greater depths, SIMS has been used to study Tiles 1, 3 and 4. SIMS uses a beam of $O_2^+$ ions of energy 5keV rastered over an area of 0.3x0.43 mm at the surface to sputter material from the sample, and positive ions are analysed in a double focusing magnetic sector spectrometer (VG Ionex IX-70S). Depth scales are derived from the sputtering time by measuring the crater depths after profiling with a Dektak 3ST: due to the substrate roughness several measurements are made and the results averaged, but the error bar is estimated as ±20%.

A number of samples have been cut from Tile 1 using a hollow drill of 20mm outside diameter (OD), which produces core samples of 17 mm OD. These samples are cut in half and polished to give sections that can be examined with a Nikon Optical Microscope. The left-hand panel of Figure 2 shows an optical micrograph from sample 10 from the horizontal part at the top of marker tile 1 (the so-called "apron") and on the right-hand side one from the bend in the front face between upper and lower parts (sample 5). In each image the metal coating is clearly visible, and it is possible to distinguish the outer W marker layer and the Mo interlayer. Sample 10 shows a film ~ 20 μm thick on top of the metallic coating; the film has a layered structure. The outer film has almost disappeared from sample 5 - in some areas just a thin film 1-2 microns thick remains - and there is no film visible on samples cut from towards the bottom of the front face.

RBS and SIMS spectra from similar points on the apron and the lower sections of the front face of tile 1 are shown in Figure 3. The $^1$H RBS spectrum from the apron shows a low concentration of W at the surface, then a gradual build-up in concentration deeper into the surface and a substantial Be peak. SIMS also shows a steady build-up in W signal into the surface, with a slight accompanying reduction in Be signal, which continues beyond 10 μm. RBS and SIMS from the upper section of the front face of tile 1, and RBS from the HFGC tile, show similar behaviour, though the depth varies. The IBA spectra have been simulated using the IBA DataFurnace NDF v9.5a [8], which inputs data from all four IBA methods ($^1$H-RBS, PIXE, $^3$He-RBS and NRA) to produce the best fitting. At the point on the apron of Tile 1 from which the RBS spectrum in Fig.3 is shown, the W concentration in the outer micron is 1%, and the total Be in the outermost $13 \times 10^{19}$ atoms cm$^{-2}$ was $9 \times 10^{19}$ atoms cm$^{-2}$. Deuterium (D, or $^1$H) was ~10% of the Be concentration, and Ni and the other elements of inconel$^©$, O and C were also present (though the sensitivity to the latter two is very low in the presence of W). Tile profiling indicates an increase in tile dimension for the upper section of the front face of Tile 1 of 10-20 microns [9] and for the HFGC tile.

A typical RBS spectrum from the lower half of the front face of Tile 1 is shown in Fig.3, and is similar to spectra from as-coated marker tiles, showing 4 μm W on a 3 μm Mo interlayer on a W substrate, but with just a very small Be surface peak indicated). The dip in the spectrum between channel numbers 200 and 500 is due to the lower scattering cross-section for Mo than for W. Concentrations of Be from NDF fitting (of all the IBA data) are of the order of $7 \times 10^{17}$ atoms cm$^{-2}$ in the lower half of the tile. The SIMS spectrum in the bottom panel of Fig.3 also shows that the Be concentration is much lower, and the W much higher, in the near-surface region shown (~6 μm) compared to the analysis from the top of the tile. There is a sharp rise in the SIMS Mo signal at a depth of ~4 microns, as expected for the marker layer.

The majority of the Be found in the inner SOL for HD-discharges appears to be restricted to the HFGC tile and parts of Tile 1. However, a significant minority (about 20%) of inner strike points in the ILW campaign were on the accessible part of Tile 4 ("corner shots"). Thus Tiles 3 and 4 have been analysed to see if Be had been deposited on these tiles during corner shots, or had migrated towards the inner divertor corner (from where samples were also analysed) as did carbon during previous JET campaigns. Tile 6 and samples from the outer divertor corner were also analysed to see whether significant deposition and/or migration was occurring in the outer divertor.

Low levels of Be were detected everywhere at the surface of tiles 4 and 6. More Be was present on the sloping parts of each tile, peaking at ~$5 \times 10^{18}$ Be atoms cm$^{-2}$ in the region that is the limit of plasma access into the corner of Tile 4, and being about $2 \times 10^{18}$ Be atoms cm$^{-2}$ in the area of Tile 4 shadowed by Tile 3. Levels on Tile 6 were similar to those at the corresponding points on Tile 4. The mean D concentration in the shadowed areas of the Tiles was ~one-half the Be level, with less on the sloping parts of the tiles. Figure 4 shows SIMS spectra from the shadowed area and the sloping part of tile 4. In the shadowed region there is clearly a more continuous Be film ~0.7 μm thick on top of the W coating, also containing D, whereas on the plasma-exposed sloping surface the Be decreases rapidly from its initial value within the first micron.

The tile 3 examined by SIMS was a "special" marker tile with just a 3 micron Mo coating on the standard W coating (i.e. no additional top layer of W). This was designed to investigate whether there was transfer of W from tile to tile at the inner divertor. The SIMS analyses of tile 3 revealed the Mo coating right from the surface. There was some W at the surface which decreased steadily into the surface, as did the Be. It is not possible to quantify the amount of Be from the SIMS spectra, however the profiles suggest there is a thinner film of variable thickness on Tile 3 than on the bottom of Tile 1 or on Tiles 4 and 6.

Among the items from the shadowed areas in the corners of the divertor that have been analysed by IBA were the inner and outer louvre clips. The louvres are a set of baffle plates that protect magnetic field coils from radiation from the plasma. Heavy deposition has been observed visually on the louvres in the past [10], and the clips are installed to allow the amount of deposition to be assessed periodically. After the ILW operations the clips were covered with a dark coloured layer like other surfaces in the shadowed regions. RBS and NRA spectra from points on the stainless steel inner and outer louvre clips that face towards the centre of the divertor are shown in Figure 5. The RBS spectrum from the outer louvre clip show sharp peaks from D and Be indicating thin films on the stainless steel substrate, but also show traces of C, O and W in the film. The RBS spectrum from the inner divertor shows broader peaks of similar size from Be, C, O and D. The W peak is also broader, and the Ni/Fe/Cr edge is shifted to lower channel number, indicating that there is a thicker film covering the stainless steel. NDF [8] simulations of the RBS and NRA spectra from the outer louvre give Be, D, C, O and W amounts of approximately 44, 27, 17, 17 and 0.9 $\times 10^{17}$ atoms cm$^{-2}$, respectively, and from the inner louvre of 28, 37, 30, 43 and $0.9 \times 10^{17}$ atoms cm$^{-2}$, respectively. The NRA spectra show similar amounts of Be at the two points, but the D level at the inner clip is greater than that at the outer clip due to retention also associated with the C.

**Discussion**

One of the objectives of the ILW is to see whether the migration of impurities to the shadowed areas mentioned in the Introduction, and the trapping of fuelling gas (which is a particular concern for ITER), is greatly reduced having removed carbon (as far as it is possible) from the plasma-facing surfaces. Encouraging signs during the 2012 operations with the ILW were that the $Z_{eff}$ for the plasma was low and there was negligible carbon in the plasma [11] Furthermore gas retention was about an order of magnitude lower for the ILW campaign than for similar pulses with the carbon wall [12].

From the results above, the overall amount of Be in the divertor, assuming an average thickness of deposit of 10 microns over the HFGC tile and the apron and top part of the front face of Tile 1, is ~25.5 cm$^{-3}$ and if an average thickness of 1 micron over tiles 3, 4 and 6 and the bottom section of Tile 1 is assumed this gives an extra ~9 cm$^{-3}$. The amount of deposition over these same areas on marker tiles exposed during the 2007-9 campaign with the carbon wall measured from cross-sections and from tile profiling was ~795 cm$^{-3}$ [13]. It is difficult to precisely compare the 2007-9 and 2011-2 campaigns since there are so many variables. However, the integrated lengths of the divertor phases for 2007-9 and 2011-2 were 33 and 13 hours, respectively. Thus the deposition of Be in 2011-2 is a factor of 9 less than the amount of carbon plus impurities found in 2007-9, allowing for the integrated length of divertor plasma phases. The difference between operations with the ILW and with the carbon wall is probably due mostly to chemistry in two ways. Firstly greatly reduced chemical sputtering in the main chamber reduces the amount of Be entering the plasma, then entering the SOL and being transported to the top of the inner divertor. Then secondly, the lack of chemical sputtering means there is negligible re-erosion of the (reduced) deposits at the inner wall, transport to the bottom of the divertor and into the shadowed regions. D/Be concentrations in films in the plasma-exposed regions are typically ~0.1, and in the shadowed regions are ~0.5-1.0. These values are similar to the values of D/C found in previous JET campaigns, and suggest that the reduction in D retention in the ILW by an order of magnitude [12] is mostly due to the reduced Be transport to the divertor, and not to reduced D retention in Be compared to C.

A surface peak of W was found on the upper front face of tile 1 which was mostly covered with Be and on tile 3 which has a Mo coating on top of the W coating. A surface peak of W was also present of the inner and outer stainless steel louvre clips ($10^{15}$ W atoms cm$^{-2}$). This is evidence that W atoms are being sputtered and re-deposited in the JET divertor. The surface peak is not evident in spectra from the HFGC tile and the apron of Tile 1, so the W atoms are not reaching the divertor from the main chamber along the inner SOL.

It is important to note that in plasma-accessible regions no area was analysed that gave total coverage of the W coated tiles with Be over the analysis areas for SIMS and RBS (0.3x0.43 mm and 1 mm diameter, respectively). The CFC substrate is rough, on a scale comparable to the film thickness. Because the plasma is incident at a low angle (<10º), asperities in the coating will face the plasma on one side and leave shadows on the other. This may be result in a "smoothing" effect of the plasma by inhibiting deposition on asperities in the W coating and enhancing deposition in shadowed areas, as was observed by Scanning Electron Microscopy (SEM) for tiles exposed in JET in 2001-4 [13]. Variation in deposit thickness may explain the presence of W right to the surface even where the mean Be film thickness is 10-20 μm, however it is also possible that there may be some inter-diffusion of Be and W. It should be noted that thin films of quite different appearance and more uniform thickness are observed in shadowed regions such as on the louvre clips, and also on (rough) divertor tiles as seen in the SIMS spectrum from Tile 4. The Be deposits will be investigated further by TFFT, using techniques such as SEM, Nuclear Microprobe, X-ray Photoelectron Spectroscopy and X-ray Diffraction [2].

**Conclusions**

Sections indicate that there are discrete films of up to 20 µm present on the apron and upper front face of Tile 1. IBA shows that the deposition region extends to the HFGC tile. The persistence of W signals in SIMS and RBS profiles right to the surface in plasma-wetted areas is probably due to erosion of Be by the plasma from small asperities in the W coating on the rough CFC substrate and re-deposition in the lee of such projections.

The total amount of Be deposition in the divertor following the first JET campaign with the ILW in 2011-2 was approximately an order of magnitude less than the C deposited in the 2007-9 campaign with the carbon wall, even allowing for the smaller integrated length of divertor phases in 2011-2 compared to the 2007-9 campaign. The Be deposition in the divertor was largely restricted to the top of Tile 1 and the HFGC tile, the region that is exposed to the inner SOL, with little migration to the corners of the divertor.

The ratios of D to Be films deposited in the divertor during 2011-2 are not dissimilar to the D to C ratios observed in previous carbon campaigns, suggesting that the reduction in D retention by an order of magnitude with the ILW compared to operations with a carbon wall [12] may be principally due to the reduction in the amount of Be deposited rather than lower retention in Be.

*This work, part-funded by the European Communities under the contract of Association between EURATOM/CCFE was carried out within the framework of the European Fusion Development Agreement. The views and opinions expressed herein do not necessarily reflect those of the European Commission.*

**Figure captions**

**Figure 1:** Areas of the JET divertor analysed, and an indication of the most common plasma cross-section during the ILW campaigns.

**Figure 2:** Optical microscope images of polished cross-sections of Tile 1: left-hand panel from the tile apron: right-hand panel from the middle of the front face

**Figure 3:** Top panel: $^1$H-RBS spectra from the apron (a) and from the lower part of the front face (b) of Tile 1 using a $^1$H beam energy of 2.3 MeV. Below: positive ion SIMS spectra from the apron (middle panel) and from the lower part of the front face (bottom panel) of the tile (5 keV $O_2^+$ beam).

**Figure 4:** Positive ion SIMS spectra from the plasma-exposed slope (left) and shadowed part (right) of the tile (5 keV $O_2^+$ beam).

**Figure 5**: $^1$H-RBS spectra (upper panel) and NRA spectra (lower panel) from the same points on the outer and inner louvre clips (a and b, respectively). Each point faces towards the centre of the divertor. The $^1$H beam energy for RBS and the $^3$He beam energy for NRA were each of 2.3 MeV. In each panel one of the spectra has been displaced vertically for clarity.

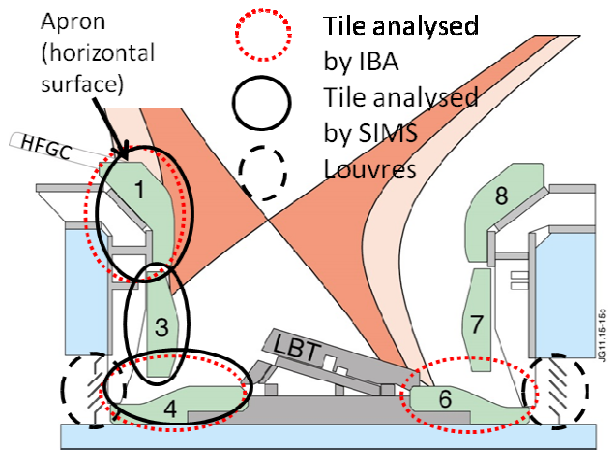

**Figure 1:** Areas of the JET divertor analysed, and an indication of the most common plasma cross-section during the ILW campaigns.

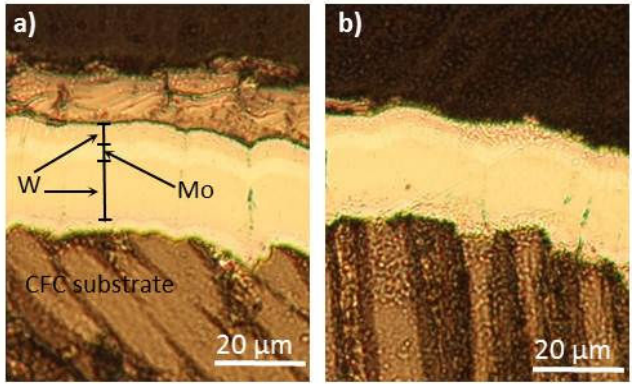

**Figure 2:** Optical microscope images of polished cross-sections of Tile 1: left-hand panel from the tile apron: right-hand panel from the middle of the front face

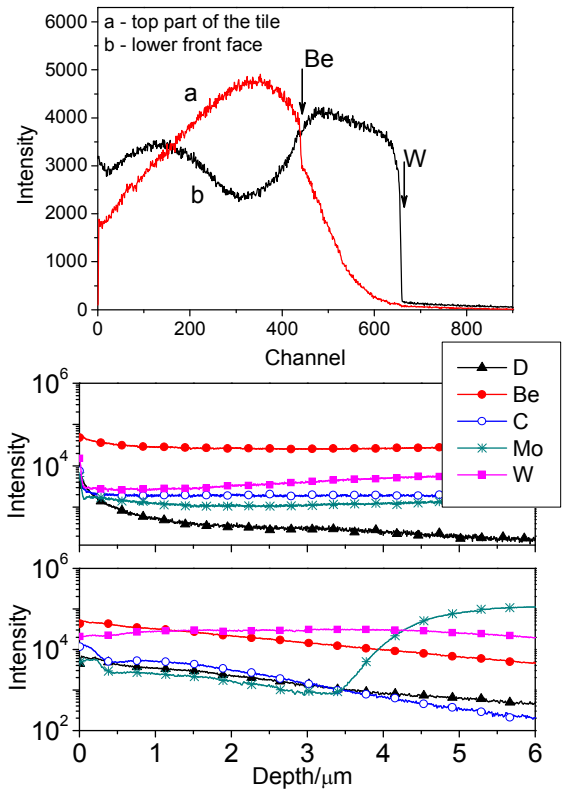

**Figure 3:** Top panel: $^1$H-RBS spectra from the apron (a) and from the lower part of the front face (b) of Tile 1 using a $^1$H beam energy of 2.3 MeV. Below: positive ion SIMS spectra from the apron (middle panel) and from the lower part of the front face (bottom panel) of the tile (5 keV $O_2^+$ beam).

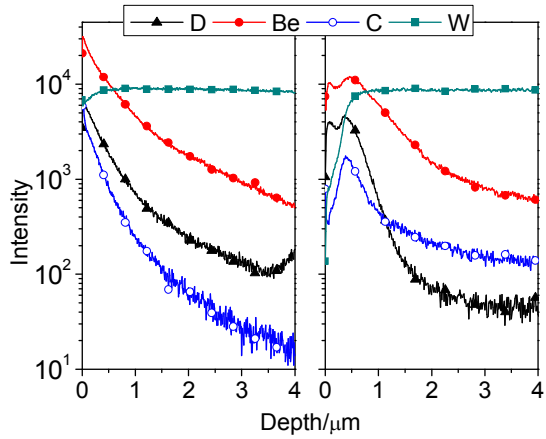

**Figure 4:** Positive ion SIMS spectra from the plasma-exposed slope (left) and shadowed part (right) of the tile (5 keV $O_2^+$ beam).

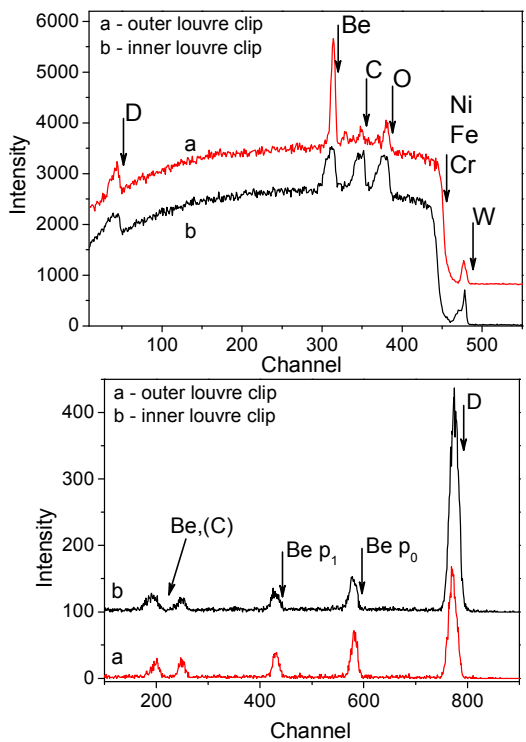

**Figure 5**: $^1$H-RBS spectra (upper panel) and NRA spectra (lower panel) from the same points on the outer and inner louvre clips (a and b, respectively). Each point faces towards the centre of the divertor. The $^1$H beam energy for RBS and the $^3$He beam energy for NRA were each of 2.3 MeV. In each panel one of the spectra has been displaced vertically for clarity.

---

[i] Corresponding address: Paul.Coad@vtt.fi